\documentclass[10pt,a4paper]{article}
\usepackage{latexsym,epsfig,multicol}
\usepackage{color}
\usepackage{graphicx}
\usepackage{pifont}
\usepackage{verbatim}

\def\beq{\begin{eqnarray}}
\def\eq{\end{eqnarray}}

\def\beqn{\begin{eqnarray*}}
\def\eqn{\end{eqnarray*}}



\usepackage{amssymb}
\usepackage{amsfonts}
\usepackage{amsmath}%
\usepackage{graphicx}

\providecommand{\U}[1]{\protect\rule{.1in}{.1in}}

\title{ Testing a possible W and Z structure at the LHC/CERN}
\author{F.~M.~L.~de Almeida Jr., H. Chavez, Y.~A.~Coutinho, \\
J.~A.~Martins Sim\~oes, A.~J.~Ramalho, S.~Wulck\\
Instituto de F\'isica, \\
Universidade Federal do Rio de Janeiro,\\
Rio de Janeiro, RJ, Brazil\\
\\
M.~A.~B.~do Vale\\
Departamento de Ci\^encias Naturais,\\
Universidade Federal de S\~ao Jo\~ao del Rei, \\
S\~ao Jo\~ao del Rei, MG, Brazil }

\date{}

\begin{document}

\maketitle 

\begin{abstract}
 
One of the first channels to be experimentally analyzed at the LHC is  $ p + p \longrightarrow l^+ + l ^- + X $. A resonance in this channel would be a clear indication of a new gauge neutral boson, as proposed in many extended models. In this paper we call attention to the possibility that the new resonance in this channel could have spin zero. A new high mass spin zero state could be a strong indication of the composite nature of the standard model particles. We have made a comparison between spin zero and spin one for the new hypothetical heavy gauge particle production and decays and we show some distributions that can easily identify their spins.

\vskip 1cm
\par
\noindent
e-mail: marroqui@if.ufrj.br, yara@if.ufrj.br, simoes@if.ufrj.br, ramalho@if.ufrj.br, steniow@if.ufrj.br, aline@ufsj.edu.br

\end{abstract}

\eject

\section{Introduction}
The Standard Model (SM) loop contributions introduce corrections at the electroweak scale that must be compensated by some nonstandard contributions. This  is the main motivation for many extended models such as supersymmetry, extra dimensions, little Higgs, technicolor and composite models \cite{Ellis:2007vz}. In composite models it is expected that more fundamental fermions, usually called preons \cite{D' Souza:1992tg}, to be responsible for the presently known matter. If this hypothesis is correct, we expect that the ordinary fermions are composite states of three preons. The Higgs field must also be a composite state and this hypothesis is expected to solve the hierarchy problem. In the same way, the SM gauge bosons could be composite states of two preons \cite{Bitar:1980fr}. In this case we must also have new states with spin zero and residual S-P interactions. In this work we present a model and predictions for this possibility to be tested at the LHC at CERN. One of t!
 he first channels for new physics at the LHC will be a new neutral heavy gauge boson $Z^\prime$ that decays into a pair of charged leptons. This same final-state channel could lead to a new heavy neutral spin zero state instead of the usual studied spin one. We present bounds on possible residual S-P interactions, distributions on new scalar states decays and a detailed comparison with similar processes for new possible spin one states. We show that once a new neutral interaction is found at the LHC, its spin can be determined.

\section{The model}
The replication of fundamental fermion representations in the SM has no deeper explanation. One possibility is that the presently known matter is the result of some more fundamental particles. But this hypothesis faces a major difficulty: despite the great progress in understanding symmetries and their breaking, we have no experimental evidence for compositeness for the presently known matter. Nevertheless, we can put forward some general points that could guide us in the search for compositeness \cite{Lopes:1987rp, Barroso:1985pu}.
\par
 There are two main points in our hypothesis that are satisfied by most models  on composite $W^{\pm}$ and $Z^0$ \cite{Bitar:1980fr}. The first very general point comes from an analogy with the quark model for hadrons: hadrons with different spins can be formed  as bound states of the same constituent quarks. So, if the presently known leptons and quarks are composite states of the fundamental preons, we expect that new spin 3/2 fermions should appear \cite{LeiteLopes:1980mh}. If we apply the same reasoning for the massive gauge bosons $ W^\pm$ and $Z^0$, we also expect  that new spin zero bound states, $\phi^\pm$ and $\phi^0$ should appear. We do not know the internal mechanism that must be present in order to form these bound states, but we do know that the SM particles follow the SM gauge group. So our second hypothesis is that the possible spin zero states also follow  the $SU(2)_L \otimes U(1)_Y$ symmetry. This means that the new scalar states must have $T=1/2$ and $Y =!
 1$ and must be in the fundamental representation
 
\begin{eqnarray}
\phi& = &\left(\begin{array}{c} \phi^+\\ \phi^0
  \end{array}\right), \qquad \tilde \phi = C \phi^*
\end{eqnarray}  
\par
These two general conditions are satisfied by most of the preon models developed so far such as Harari-Shupe \cite{Harari:1979gi,Shupe:1979fv} and Fritzsch-Mandelbaum \cite{Fritzsch:1981tt} models. 
Leptons and quarks must be in the same SM basic representation and if we add right-handed neutrino singlets the new interaction between scalars and fermions is:

\begin{equation}
{\cal L}_{fermions} = g_1 \left[ \bar L \, \phi \, e_R + \bar e_R \, \phi ^{\dagger} \,  L \right] + g_2 \left[ \bar L \, \tilde \phi \,  \nu_R + \bar \nu_R \,  \tilde \phi^{\dagger} \, L \right]
\end{equation}

This interaction can be rewritten as 

\begin{equation}
{\cal L}=g_{1}\left[  \bar \nu_{eL}^{\, \prime} e_R \phi^ + + \bar e_R %
\nu_{eL}^{\, \prime} \phi^- + \bar e e \phi^0 \right]  + g_2 \left[
-\bar \nu_{eR}^{\, \prime} e_L \phi^+ - \bar e_L \nu_{eR}^{\, \prime}%
\phi^- + \bar \nu_e  \nu_e \phi^0 \right] 
\end{equation}

In this interaction, there are two new coupling constants $g_1$ and $g_2$ and the scalar state masses must be experimentally determined. The generalization to other leptonic families and quark sectors is straightforward. 
The three scalar states $\phi^\pm$ and $\phi^0$ must all have the same mass. We can add a fourth neutral scalar singlet $\chi^0$. Through $\chi^0$, $\phi^0$ mixing one can  generate a different mass between neutral and charged states. Let us call $H$ the Standard Model Higgs field that develops
a non-zero vacuum expectation value at the Fermi scale. Then the invariant mass lagrangian is given by:

\begin{eqnarray}
 &&{\cal L}_{mass}=m^2_{\phi} \phi^+ \phi +  m^2_{\chi} \chi^0 \chi^0+  \mu^2 H^+ H  
 -\lambda(H^+ H)^2 - \nonumber \\ 
&& \lambda_2 ( H^+ H )(\phi^+ \phi) - \lambda_3 (H^+ H )(\chi^0 \chi^0) + \lambda_1(\phi^+ H) \chi^0 + ...
\end{eqnarray}

Other quartic terms can be added but they will be not important for the new scalar mass spectrum, at tree level. 
The last term in this equation will mix the $\phi^0$ and $\chi^0$ fields and will give diferent $\phi^+ $ and $\phi^0$ masses. In terms of the physical $ M_{\phi^0},M{\chi^0},m_{\phi^\pm}$ masses  the $\phi^0$,$\chi^0$ mixing angle is given by:

\begin{equation}
\cos {2 \, \alpha}=\frac{M^2_{\phi^0} +M^2_{\chi^0} -2m_{\phi^\pm}^2}{M^2_{\phi^0} - M^2_{\chi^0}}
\end{equation}

\par
An essential point in our model is that the new scalar fields do not develop non-zero vacuum expectation values as the SM Higgs particles and no additional hypothesis as discrete symmetries \cite{Ma:2008uz} are introduced. Our model will be also quite diferent from two-Higgs models \cite{Gunion:2002zf}.
\par
Another important consequence of the model is that the scalar state $\chi^0$ can have a low mass and it is a candidate for Dark Matter. The possibility of a spin-zero candidate was studied by Fayet \cite{Fayet:2004bw} and more recently by Barger, Keung and Shaughnessy \cite{Barger:2008qd}. The relevance of spin-zero Dark Matter for the LHC searches was analised by  Baer and Tata \cite{Baer:2008uu}.
\par
 The $SU(2)_L \otimes U(1)_Y$ invariant interaction between scalars and gauge bosons can be obtained from:

\begin{equation}
{\cal L}_{gauge}(\phi) = \left[ \left( i \,\,\partial_{\mu} - g \frac{\tau_a}{2} \, A_{\mu}^a - \frac{g^{\, \prime}}{2} \, Y_{\phi} B_{\mu} \right)
 \phi  \right]^2
\end{equation}

In order to compare our results for a spin zero state with possible new neutral gauge bosons of spin one, we have considered a general fermion-fermion coupling with a new $Z^\prime$ given by:

\begin{eqnarray}
{\cal L}_{NC} = -\frac{g}{2\sin\theta_W}\bar \psi_i \gamma^{\mu}(g_V - g_A\gamma_5) \psi_i {Z^{\prime}}^{\mu }
\end{eqnarray}
These new couplings are normalized as in spin zero interactions and the new spin one mass is taken equal as in the spin zero case.

\section{Bounds on S-P interactions}

As we have no direct experimental evidence for new particles and interactions, we expect that the scalar masses are  to be placed in the region above a few hundred GeV. Our new fermion-scalar interaction generates at low energy  new residual effective S-P interactions. Bounds on these new interactions from muon decay were studied in \cite{Lopes:1987rp}. With the latest experimental values on muon decay parameters \cite{PDG} we have updated these results and found with $95 \% $ C.L.:

\par
\begin{eqnarray}
&& 0.0 < \vert g^S_{RR} \vert < 7.75 \times 10^{ -2} \\ \nonumber
&& 0.0 < \vert g^S_{LR} \vert < 7.75 \times 10^{ -2}\\ \nonumber
&& 0.0 < \vert g^S_{RL} \vert < 0.41 
\end{eqnarray}
\par
where
\begin{equation}
g^{S}_{a,b}=\frac{\sqrt 2}{G_F}\frac{g _i g_j}{m^2_{\phi}}
\end{equation}
with a, b = R, L and i, j = 1, 2. Other bounds \cite{Lopes:1987rp} can be obtained from the $ K^0 - \bar K^0 $ system but they do not improve the above bounds. There are experimental bounds on new neutral scalar particles \cite{PDG} that imply a lower bound on their masses of a few hundred GeV's. This means that the bounds on $g_1$ and $g_2$ from the above relations are not very stringent.
 
In order to set more restrictive bounds on the fermion-scalar boson couplings $g_1$ and $g_2$, we carried out a $\chi^2$ analysis, based on the experimental data collected by the OPAL Collaboration for the quark anti-quark production at a center-of-mass energy $\sqrt{s} = 189$ GeV \cite{OPA}. A $\chi^2$ estimator was defined as
\begin{equation}
\chi^2(g_1,g_2,M_{\phi^0}) \equiv \sum_{i=1}^{N_b} \left (\frac {X_i - X_i^{exp}}{\Delta X_i^{exp}} \right )^2
\end{equation}
where $X_i^{exp} = d\sigma^{exp}/d \, \vert \cos \theta_i\vert$ stands for the experimental value of the angular distribution in the $i^{th}$ bin and $\Delta X_i^{exp}$ the corresponding experimental uncertainty, which includes both the statistical and systematic errors. The term $X_i = d\sigma /d \, \vert \cos\theta_i \vert$ denotes the theoretical prediction for the angular distribution in the $i^{th}$ bin, taking into account the scalar interactions. The sum runs over $N_b = 20$ equal-size bins. To make the comparison with OPAL data more meaningful, our simulation selected a high-energy hadronic sample for which the effective center-of-mass energy satisfies $\sqrt{s^\prime} > 0.85 \sqrt{s}$. The modeling of initial-state radiation followed the structure function approach of ref. \cite{SKR}. Figure 1  shows the one- and two-degree of freedom $95\%$ C. L. exclusion contours in the $g_1-g_2$ plane, both for $M_{\phi^0} = 500$ GeV and $1$ TeV. The one-degree of freedom contou!
 rs were derived by cutting at $\Delta \chi^2 = \chi^2- \chi^2_{min} = 3.84$, whereas the correlated limit contours were obtained by cutting the $\chi^2$ estimator $5.99$ units above the minimum.
In numerical estimates throughout this paper we have considered typical values $ g_1 = 0.2$ and $ g_2 = 0.3$.

\begin{figure} 
\begin{center}
\includegraphics[width=.6\textwidth]{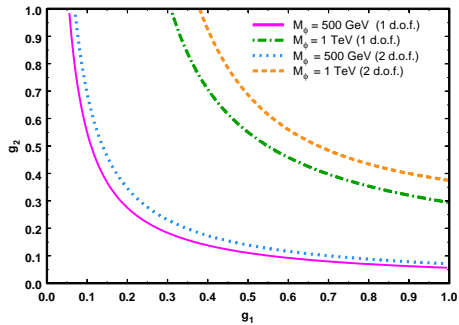}
\caption{The exclusion contours with one- and two-degree of freedom $95\%$ C. L. in the $g_1-g_2$ plane for $M_{\phi^0} = 500$ GeV and $M_{\phi^0} = 1$ TeV.}
\end{center} 
\end{figure}

\section{New scalar particles at the LHC}

The new scalar states $\phi^\pm$ and $\phi^0$ can be singly produced at the LHC and detected through their decay into pairs of standard fermions. The charged scalars $\phi^\pm$ can be detected in two jets final states. One of the first and cleanest channels to be experimentally studied at the LHC will be $ p + p \longrightarrow l^- + l ^+ + X $, where $l= e, \mu $. There are many theoretical models beyond the SM that predict a new neutral gauge boson. In the present model, the neutral scalar field $\phi^0$ can also contribute to this process. If a new resonance is founded in this channel at the LHC, one of the first points to be investigated is its intrinsic angular momentum. In this section we present a detailed comparison of distributions of final charged leptons (muons or electrons) for both spin one and spin zero. Besides the interaction Lagrangian presented in Section 2 we have considered a generic new $Z^\prime$ interacting with charged leptons as given by Eq. 5, with !
 V-A couplings.

We applied the following cuts on the final fermions: $\vert\eta\vert \leq 2,5$ since the LHC detectors have better tracking resolution in this $\eta$ range and $M_{\mu^- \mu^+} > 400$ GeV to avoid SM background in special from the $Z^0$ peak. The model based on the Lagrangians (Eqs. 3, 4 and 5) was implemented in Comphep package \cite{COM}, allowing the generation of events and the calculations of total cross sections and distributions.

The total cross section for $ p + p \longrightarrow l^+ + l ^- + X $ is shown in Figure 2. The signal is above the SM background for masses up to $2$ TeV. Higher masses will require high luminosity and this background can be reduced by applying stronger cuts.

\begin{figure} 
\begin{center}
\includegraphics[width=.6\textwidth]{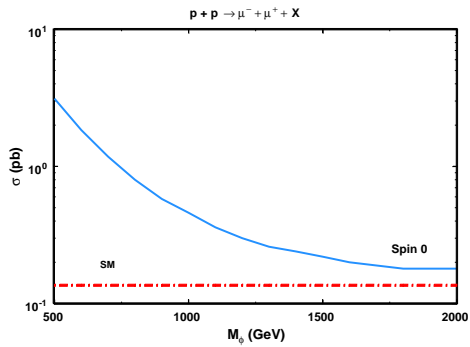}
\caption{Total cross section as a function of $M_{\phi^0}$ for the process $ p + p \longrightarrow \mu^- + \mu ^+ + X$ considering the spin zero model and the standard model.}
\end{center} 
\end{figure}

In the Figures 3 and 4 we show the $\eta_{\mu^{-}}$ normalized distributions for two $\phi^0$ different mass values. This is a very interesting variable since it depends only on angular determination of the final charged leptons. There is a clear difference for spin zero and spin one resonances. It can be noted that in the spin one case, the final leptons are produced with higher $\vert \eta \vert $ than in the spin zero case.

\begin{figure} 
\begin{center}
\includegraphics[width=.6\textwidth]{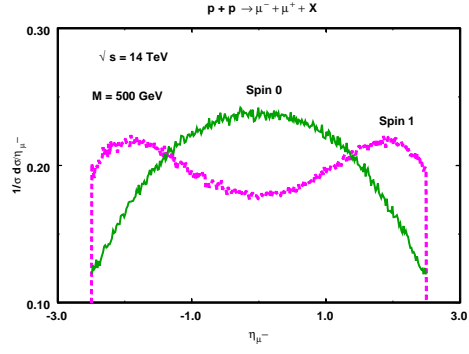}
\caption{Normalized $\eta_{\mu^{-}}$ distribution for $M_{\phi^0}=M_{Z^\prime}= 500$ GeV in the process $ p + p \longrightarrow \mu^- + \mu ^+ + X$.}
\end{center} 
\end{figure}

\begin{figure} 
\begin{center}
\includegraphics[width=.6\textwidth]{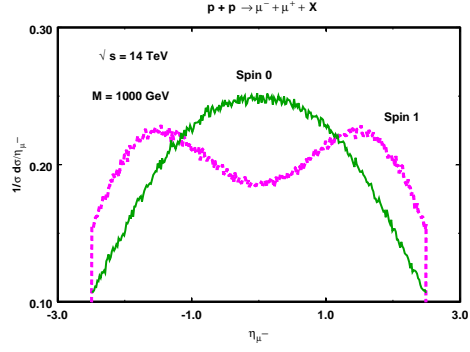}
\caption{Normalized $\eta_{\mu^{-}}$ distribution for $M_{\phi^0}=M_{Z^\prime}= 1000$ GeV in the process $ p + p \longrightarrow \mu^- + \mu ^+ + X$.}
\end{center} 
\end{figure}

 Another useful variable that also shows spin one and spin zero different distributions is the cosine of the angle in the $CM$ of the final leptons between one of the final state charged leptons along the direction of their boost. This is shown in Figures 5 and 6. The clear differences between spin zero and spin one distributions are maintained for $M_{\phi^0}= 500$ GeV and $M_{\phi^0}= 1000$ GeV. 

\begin{figure} 
\begin{center}
\includegraphics[width=.6\textwidth]{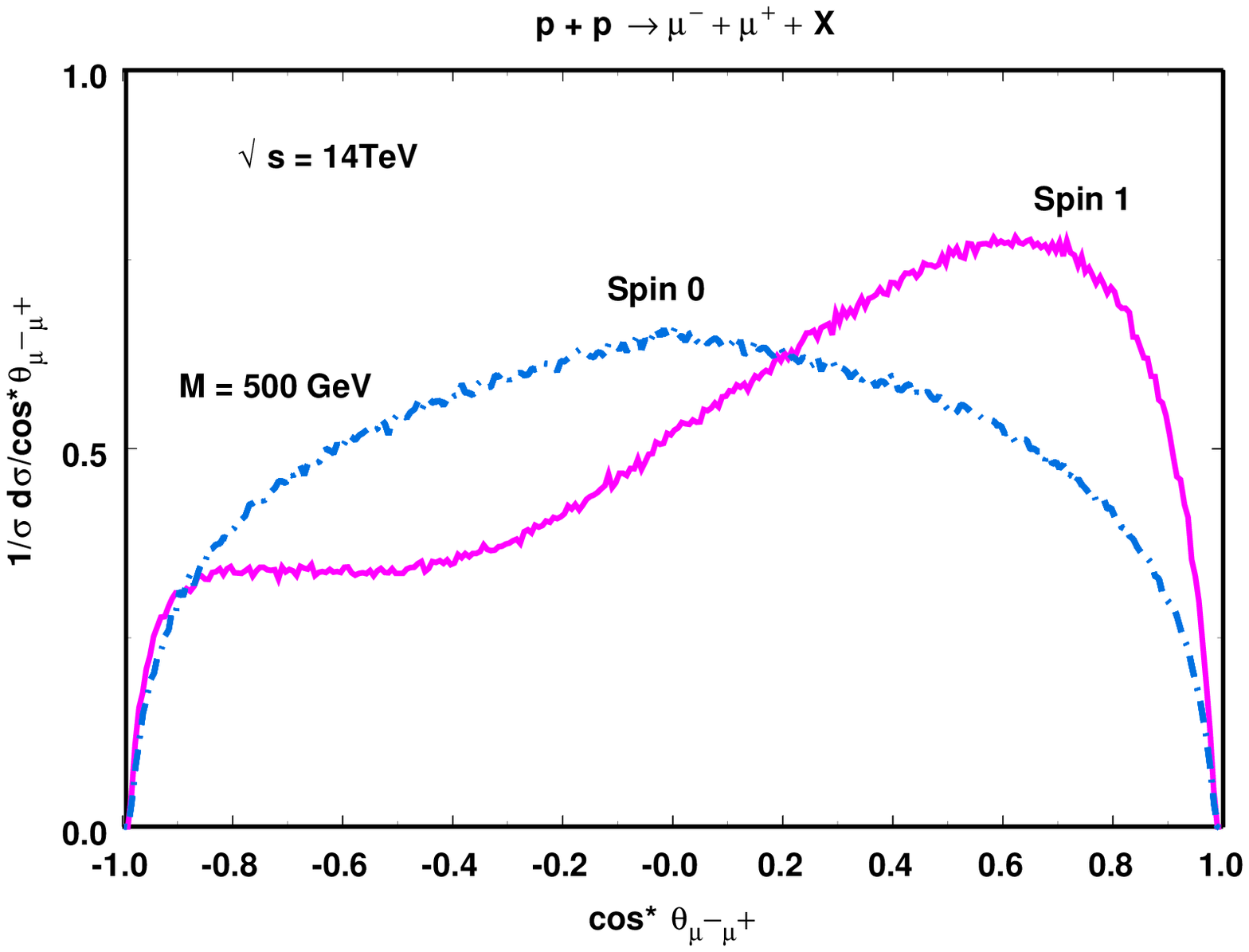}
\caption{Normalized $cos{\,\theta_{\mu^- \mu^+}^*}  $ angular distribution  for $M_{\phi^0}=M_{Z^\prime}= 500$ GeV in the process $ p + p \longrightarrow \mu^- + \mu ^+ + X$.}
\end{center} 
\end{figure}

\begin{figure} 
\begin{center}
\includegraphics[width=.6\textwidth]{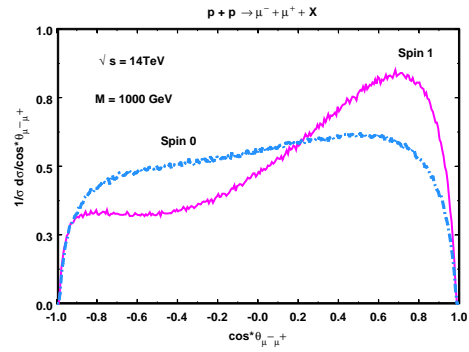}
\caption{Normalized $cos{\,\theta_{\mu^- \mu^+}^*} $ distribution  for $M_{\phi^0}=M_{Z^\prime}= 1000$ GeV in the process $ p + p \longrightarrow \mu^- + \mu ^+ + X$.}
\end{center} 
\end{figure}

\eject

\section{Neutrino masses}
\par
As neutrinos have non-zero masses, the inclusion of right-handed neutrino \break singlets seems to be a natural scenario  to account for their masses\cite{Schechter:1980gr}. However, as these singlets are completely neutral ($Q = 0$; $T = 0$; $Y = 0 $) they have no interactions with the gauge vector bosons and some additional hypothesis must be done to give non-zero masses. One possibility is the see-saw mechanism with Majorana mass terms. But this mechanism implies very high new neutrino masses that must be avoided  by some new symmetry. In the present model, the new charged scalars can contribute as shown in Figure 7, and  we can have Dirac and or Majorana mass terms.

\begin{figure} 
\begin{center}
\includegraphics[width=.6\textwidth]{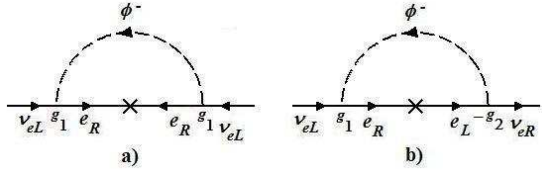}
\caption{One-loop contribution to neutrino masses.}
\end{center} 
\end{figure}

The dominant contribution to the neutrino masses is given by

\begin{equation}
M_{\nu} \simeq \frac{g_{i}g_{j}}{4\pi^2}\frac{m_e ^3}{m_{\phi}^2}\, \ln(\frac{m_{\phi}^2}{m_e ^2})
\end{equation}
where $i= j= 1$ for Dirac masses and $i= 1$, $j= 2$ for Majorana masses. 

 These results suggests that neutrino masses must follow the charged leptons mass spectrum. A very important aspect of this model is that it is not necessary to have extremely high Majorana mass neutrinos as in the see-saw model. Then we can expect that Majorana neutrinos could have lower masses \cite{Almeida:2000pz} and can be searched at the LHC. But we have not included neutrino mixings and new neutral scalar contributions. The inclusion of more parameters can, in principle, modify this simple neutrino mass spectrum and will be discussed in detail in a separate future paper. 

\section{Conclusion}

In this paper we have presented a study of a possible composite model for the electroweak gauge bosons 
$W^{\pm}$ and $Z^0$. The near start of the experimental program of the LHC will allow to test this hypothesis. 
New bounds on S-P interactions were obtained from the OPAL data on the angular quark anti-quark distributions produced 
in $e^+ + e^-$ collisions.
There is a very clear signature for new spin zero particles in $ p + p \longrightarrow l^+ + l ^- + X $. 
We have done a detailed comparison with a possible spin one new gauge boson and show that a clear spin determination \cite{Wang:2008sw} can be done in the $\eta_{l^{-}}$ and  $\cos^* \theta_{\mu^- \mu^+} $ distributions. The new scalar states $ \phi^0$ and $\phi^{\pm}$ can give important contributions to neutrino masses to both Dirac and Majorana terms.

\vskip 5cm

\textit{Acknowledgments:} 
This work was supported by the following Brazilian agencies: CNPq, FAPERJ and FAPEMIG.


\begin{thebibliography}{99}
\bibitem{Ellis:2007vz} J.~Ellis, AIP Conf.\ Proc.\  {\bf 957}, 38 (2007). [arXiv:hep-ph/0710.0777].
\bibitem{D' Souza:1992tg} I.~A.~D' Souza and C.~S.~Kalman, {\it  Singapore, Singapore: World Scientific (1992) 108 p}.
\bibitem{Bitar:1980fr} K.~M.~Bitar, Phys.\ Rev.\  D {\bf 24}, 2654 (1981);
N.~S.~Craigie and J.~Stern, Fortsch.\ Phys.\  {\bf 34}, 261 (1986); 
P.~D.~B.~Collins and N.~A.~Speirs, J.\ Phys.\ G {\bf 11}, L115 (1985); 
M.~Yasue, Phys.\ Rev.\  D {\bf 39}, 3458 (1989);
U.~Baur, D.~Schildknecht and K.~H.~G.~Schwarzer, Phys.\ Rev.\  D {\bf 35}, 297 (1987); 
Y.~V.~Novozhilov, Phys.\ Lett.\  B {\bf 225}, 165 (1989); 
R.~Casalbuoni, S.~De Curtis, D.~Dominici and R.~Gatto, Phys.\ Lett.\  B {\bf 155}, 95 (1985); 
A.~Queijeiro and D.~M.~Tun, Z.\ Phys.\ C {\bf 47}, 63 (1990); 
A.~Martinez, A.~Queijeiro and D.~M.~Tun, Phys.\ Scripta {\bf 45}, 425 (1992); 
K.~Sugita, Y.~Okamoto and M.~Sekine, Int.\ J.\ Theor.\ Phys.\  {\bf 36}, 1309 (1997); 
A.~Doff, A.~A.~Natale and P.~S.~Rodrigues da Silva, Phys.\ Rev.\  D {\bf 77}, 075012 (2008).
\bibitem{Lopes:1987rp} J.~H.~Lopes and J.~A.~Martins Sim\~oes, Phys.\ Rev.\  D {\bf 35}, 3428 (1987).
\bibitem{Barroso:1985pu} M.~Barroso, M.~E.~Magalh\~aes, J.~A.~Martins Sim\~oes, N.~Fleury, and J.~Leite Lopes,
Phys.\ Rev.\  D {\bf 34}, 1451 (1986).
\bibitem{LeiteLopes:1980mh} J.~Leite Lopes, J.~A.~Martins Sim\~oes and D.~Spehler, Phys.\ Lett.\  B {\bf 94}, 367 (1980); J.~Leite Lopes, D.~Spehler and J.~A.~Martins Sim\~oes, Phys.\ Rev.\  D {\bf 25}, 1854 (1982); 
C.~J.~C.~Burges and H.~J.~Schnitzer, Nucl.\ Phys.\  B {\bf 228}, 464 (1983); 
Y.~Tosa and R.~E.~Marshak, Phys.\ Rev.\  D {\bf 32}, 774 (1985); 
O.~Cakir and A.~Ozansoy, Phys.\ Rev.\  D {\bf 77}, 035002 (2008).
\bibitem{Harari:1979gi} H.~Harari, Phys.\ Lett.\  B {\bf 86}, 83 (1979).
\bibitem{Shupe:1979fv} M.~A.~Shupe, Phys.\ Lett.\ B {\bf 86} (1979) 87.
\bibitem{Fritzsch:1981tt} H.~Fritzsch and G.~Mandelbaum, Phys.\ Lett.\  B {\bf 109} (1982) 224.
\bibitem{Ma:2008uz} E.~Ma, Mod.\ Phys.\ Lett.\  A {\bf 23}, 647 (2008).
[arXiv:hep-ph/0802.2917].
\bibitem{Gunion:2002zf} J.~F.~Gunion and H.~E.~Haber, Phys.\ Rev.\  D {\bf 67}, 075019 (2003).
  [arXiv:hep-ph/0207010], J.~F.~Gunion, H.~E.~Haber, G.~L.~Kane, and S.~Dawson, The Higgs Hunter's Guide (Perseus Publishing, Cambridge, MA, 1990).
\bibitem{Fayet:2004bw} P.~Fayet, Phys.\ Rev.\  D {\bf 70}, 023514 (2004). [arXiv:hep-ph/0403226].
\bibitem{Barger:2008qd} V.~Barger, W.~Y.~Keung and G.~Shaughnessy, Phys.\ Rev.\  D {\bf 78} (2008) 056007 [arXiv:hep-ph/0806.1962].
\bibitem{Baer:2008uu} H.~Baer and X.~Tata, [arXiv:hep-ph/0805.1905].
\bibitem{PDG} C.~Amsler {\it et al.}, (Particle Data Group), Physics Letters B {\bf 667}, 1 (2008). 
\bibitem{OPA} OPAL Collaboration, Eur. Phys. J. C {\bf 13},553 (2000), CERN-EP/99-097.
\bibitem{SKR} M.~Skrzypek and J.~Jadach, Z. Phys. C {\bf 49}, 577 (1991).
\bibitem{Schechter:1980gr} J.~Schechter and J.~W.~F.~Valle, Phys.\ Rev.\  D {\bf 22}, 2227 (1980).
\bibitem{Almeida:2000pz} F.~M.~L.~Almeida, Y.~A.~Coutinho, J.~A.~Martins Sim\~oes and M.~A.~B.~do Vale, Phys.\ Rev.\  D {\bf 62}, 075004 (2000).
\bibitem{COM} E.~Boos {\it et al.}  [CompHEP Collaboration],
Nucl.\ Instrum.\ Meth.\  A {\bf 534}, 250 (2004). [arXiv:hep-ph/0403113].
\bibitem{Wang:2008sw}  L.~T.~Wang and I.~Yavin. [arXiv:hep-ph/0802.2726].
 
\end{thebibliography}
\end{document}